\documentclass[twocolumn, trackchanges]{aastex701}

\usepackage{amsmath} 
\usepackage{tabularx}
\usepackage{makecell}
\usepackage{graphicx} 
\usepackage{array}
\usepackage[none]{hyphenat} 

\begin{document}

\title{Joint Multiband Photometry with \texttt{crowdsource}}

\author[0009-0002-2434-5903]{Jayashree Behera}
\affiliation{Space Telescope Science Institute, 3700 San Martin Drive
Baltimore, MD 21218}
\email[show]{jayashreeb@stsci.edu}  

\author[0000-0002-3569-7421]{Edward F. Schlafly} 
\affiliation{Space Telescope Science Institute, 3700 San Martin Drive
Baltimore, MD 21218}
\email{eschlafly@stsci.edu}

\author[0000-0002-1125-7384]{Aaron M. Meisner}
\affiliation{NSF National Optical-Infrared Astronomy Research Laboratory, 950 N. Cherry Ave., Tucson, AZ 85719, USA}
\email{aaron.meisner@noirlab.edu}

\author[0000-0002-5166-8671]{Lucas Napolitano}
\email{lucas.napolitano@noirlab.edu}
\affiliation{NSF National Optical-Infrared Astronomy Research Laboratory, 950 N. Cherry Ave., Tucson, AZ 85719, USA}

\begin{abstract}
We present a new multiband extension to the \texttt{crowdsource} photometric pipeline, enabling simultaneous fitting across multiple imaging bands in crowded fields.  The core idea is that multiple images of the same part of the sky should have the same sources at the same locations; only the fluxes in the different images should be allowed to vary in fitting.  The framework also allows us to use all images of a given region to detect faint sources, with configurable weighting among the different bandpasses as appropriate for different source spectra.  Similar concepts are already present in other crowded field packages like DAOPHOT and DOLPHOT; we now include it in the \texttt{crowdsource} fitting approach.  We describe the mathematical formulation of the multiband fit and demonstrate its performance using the \textit{Wide-field Infrared Survey Explorer} (WISE) W1 and W2 imaging as a concrete application. The multiband algorithm improves flux consistency and reduces band-to-band positional scatter relative to independent-band fitting. We test the method on unWISE coadded tiles spanning both sparse and crowded regions and quantify improvements in photometric agreement and astrometric stability. This framework provides a general foundation for future multiband crowded-field catalogs.
\end{abstract}



\section{Introduction}
\label{sec:intro}
Large-area infrared imaging surveys such as the \emph{Wide-field Infrared Survey Explorer} (WISE) \citep{2010AJ....140.1868W} provide an unprecedented view of the sky at infrared wavelengths, enabling a wide range of astrophysical applications that rely on consistent photometry across large angular scales. Extracting accurate photometry from these data is challenging due to source crowding, blending, spatially varying backgrounds, the relatively large pixel scale of WISE/unWISE imaging (2.75$''$ per pixel), and the broad point-spread function (PSF) characteristic of infrared survey imaging relative to optical imaging. These challenges motivate photometric methods that explicitly model overlapping sources while remaining robust and efficient, so as to enable application to large, survey-scale datasets.

Crowded-field photometry has a long history in optical and infrared astronomy, with widely used pipelines such as \texttt{DAOPHOT} \citep{1987PASP...99..191S} and \texttt{DOLPHOT} \citep{2000PASP..112.1383D, 2016ascl.soft08013D} designed to perform point-spread function (PSF)-based fitting in dense fields. These tools have been applied to many images of crowded fields.  However, these tools frequently require a significant amount of tuning to deliver good photometry in different environments, and employ a somewhat bespoke optimization approach rather than relying on off-the-shelf optimization tools.

The \texttt{crowdsource} pipeline \citep{2018ApJS..234...39S, 2019ApJS..240...30S} was developed to take advantage of large sparse linear algebra packages to fit for large numbers of stars in crowded fields, with an emphasis on robustness and computational efficiency. The code is implemented in Python and is publicly available,\footnote{\url{https://github.com/schlafly/crowdsource}}. The method explicitly models overlapping sources together with a smooth, parametric sky background and employs a linearized fitting framework that scales efficiently to images containing large numbers of sources. Its single-band implementation on the DECam
Galactic Plane Survey (DECaPS) \citep{2018ApJS..234...39S} and WISE \citep{ 2019ApJS..240...30S} demonstrated that this approach can deliver uniform and reliable photometry across the full sky, including in regions of moderate to high source density.

The \texttt{crowdsource} pipeline, as applied to DECaPS and WISE, is missing one feature common to DAOPHOT and DOLPHOT.  It fits individual images in isolation, obtaining different sets of sources at different positions in each image.  It does not enforce any cross-band consistency. When applied to multi-band survey data, this leads to important differences in source positions and fluxes between bands. These inconsistencies propagate into derived colors, source classifications, proper motions, and other downstream measurements that depend on consistent cross-band photometry and astrometry.

A variety of multiband photometric strategies have been explored to address this limitation, including forced photometry in which source positions are fixed from a reference band \citep{2016AJ....151...36L}. While effective in many contexts, such approaches can inherit biases from the chosen reference data or fail to fully exploit correlated information across bands, particularly when signal-to-noise ratios differ significantly between filters or when blending varies with wavelength. Other approaches construct a combined detection image from multiple bands, as implemented in pipelines such as \texttt{DAOPHOT} and \texttt{DOLPHOT}. However, such methods do not explicitly account for band-dependent PSFs in the detection step. In contrast, our approach performs detection using a multiband significance image that incorporates the appropriate PSF and noise properties of each band, similar in spirit to matched-filter approaches for multi-image detection \citep{1999AJ....117...68S}.

In this work, we introduce a multiband extension to the \texttt{crowdsource} pipeline that fits a simultaneous, self consistent model across multiple imaging bands. The algorithm enforces shared source positions while solving for band-specific fluxes using band-dependent PSF models, with all parameters optimized within a unified linearized least-squares framework. By combining the signal from all bands in a single joint solution, the multiband formulation leads to improved source positions, deblending, fluxes, and colors compared to independent single-band fits, particularly when one band is shallower or noisier. This design preserves the computational efficiency and robustness of the original pipeline while naturally incorporating constraints from multiple images.

This paper presents the design, implementation, and validation of the multiband \texttt{crowdsource} algorithm, using WISE W1 and W2 imaging as a concrete application and testbed. Our focus is on the algorithmic formulation and code-level performance, including photometric consistency, astrometric stability, and residual behavior. We intend to apply this pipeline to the full-depth, multiband coadds from the WISE mission in a coming paper.

\section{Algorithm}
\label{sec:algo}

Our multiband extension of \texttt{crowdsource} retains the core philosophy and overall structure of the single-band pipeline described in \citet{2019ApJS..240...30S}. As in the original formulation, the image is modeled as a sum of PSF-shaped point sources whose positions and fluxes are refined iteratively through a linear least-squares solution. 
Starting from an initial model (typically zero sources and an initial PSF estimate for each band), the pipeline proceeds iteratively. Each iteration consists of the following steps:

\begin{enumerate}
    \item \textbf{Sky estimation:} A smooth background is estimated from the current residual image and subtracted.
    
    \item \textbf{Source detection:} Candidate sources are identified as peaks in a matched-filter significance image.
    
    \item \textbf{Linear fit:} The fluxes, positions, and sky parameters are solved simultaneously using a linearized least-squares system.
    
    \item \textbf{Position refinement:} Source positions are updated iteratively using derivative PSF components within the linearized fit, allowing any shifts to be solved simultaneously with fluxes while accounting for neighboring sources in the model.

    
    \item \textbf{PSF update:} The point-spread function model is refined using bright, well-measured sources.
\end{enumerate}

These steps are repeated, allowing progressively fainter and more blended sources to be identified and incorporated into the model. The iteration continues until convergence, after which a final pass is performed with fixed source positions to obtain stable flux measurements. Further details of the pipeline and its components are provided in Appendix \ref{sec:appendix_pipeline}

The multiband extension follows the same overall structure, with the key modification that all available bands are fit \emph{simultaneously}, enforcing a shared set of source positions while solving for band-specific fluxes using band-dependent point-spread functions and noise models. The other major improvement is that source detection is likewise generalized to operate on a multiband significance image, enabling \texttt{crowdsource} to optimally detect point sources of a given SED across the set of available images, using appropriate point spread functions for each. Details of our modifications are described in the following sections.

\begin{figure*}
    \centering
    \includegraphics[width=\textwidth]{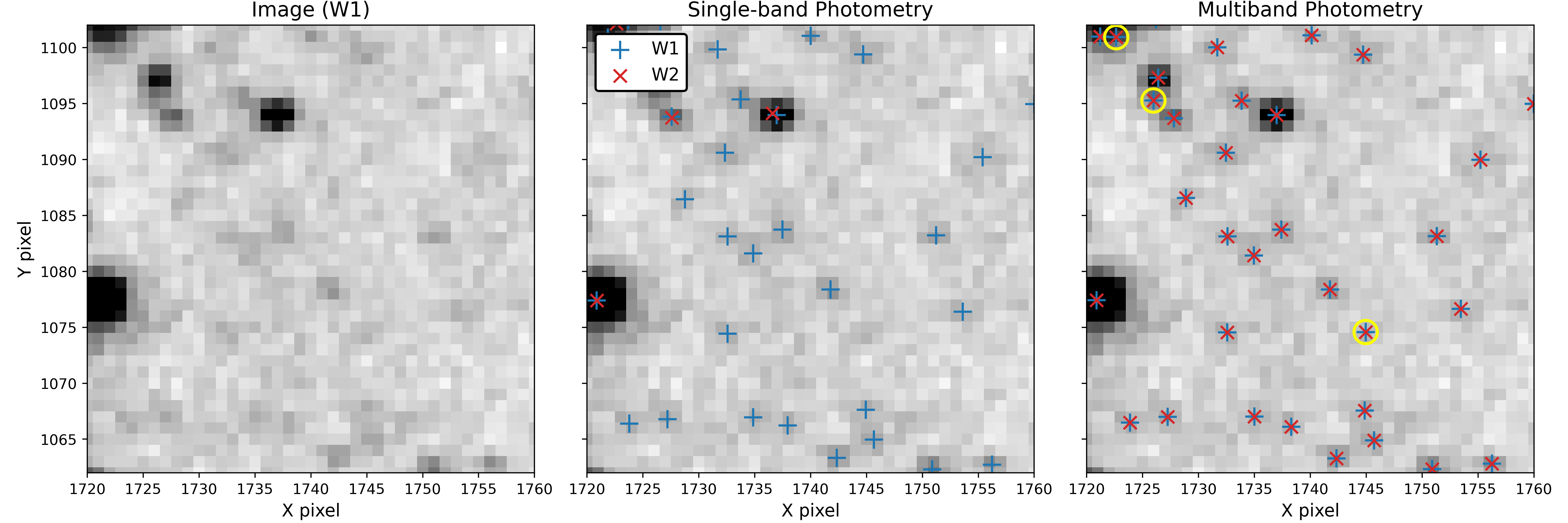}
    \caption{
    Illustration of single-band versus multiband source catalog in a representative high-latitude unWISE field (\texttt{1497p015}). 
    \textit{Left:} W1 image cutout.
    \textit{Middle:} Source positions recovered from independent single-band fits in W1 (blue) and W2 (red), showing band-dependent centroid scatter and missed faint detections in W2.
    \textit{Right:} Source positions recovered from the multiband fit, yielding a single, self-consistent set of centroids across bands. The multiband fit recovers all W2 sources associated with single-band W1 detections and additionally identifies new faint sources (highlighted with yellow circles) that are not detected in either single-band catalog. 
    }
    \label{fig:algo_spoiler}
\end{figure*}

Figure~\ref{fig:algo_spoiler} illustrates three generic outcomes of the joint formulation: (i) a single, self-consistent set of source positions across bands; (ii) improved recovery of sources that are faint in one band, particularly the noisier band, via joint source modeling; and (iii) the detection of additional faint sources that are individually marginal in both bands but become significant when modeled jointly. Although demonstrated here for WISE W1 and W2, the algorithm is defined independently of this and these outcomes reflect general properties of the joint fitting formulation and are not specific to this particular dataset.

\subsection{Significance Images and Joint Detection}
\label{sec:sigimage}

The \texttt{crowdsource} package detects sources using a matched-filter ``significance image'' for each band. Its goal is to answer a simple question at every pixel: ``If a point source were located here, how significant would it be above the local noise and background?''

For band $b$ with background–subtracted image $I_b$ and noise map $\sigma_b$,  the significance image is
\begin{eqnarray}
S_b(x,y) &=& \frac{N_b(x,y)}{\sqrt{V_b(x,y)}}, \label{eq:sb}\\
where, \nonumber\\
N_b(x,y) &=& \Big[\, I_b\,\frac{1}{\sigma_b^2} \;\star\; P_b \,\Big](x,y), \nonumber\\
V_b(x,y) &=& \Big[\, \frac{1}{\sigma_b^2} \;\star\; P_b^2 \,\Big](x,y). \nonumber
\end{eqnarray}

where $\star$ denotes convolution and $P_b$ is the band PSF (normalized to unit flux). The numerator $(P_b \star I_b)$ measures  the amount of flux entering a PSF-weighted aperture centered at $(x,y)$, weighting by the uncertainties. The denominator quantifies the expected root-mean-square uncertainty in this matched-filtered quantity. The peaks of $S_b$ therefore correspond to likely source positions.

For joint detection, we combine the individual-band matched-filter numerators and variances using optional band weights $w_b$ (normalized to 1). \\
\begin{align}
N_{\rm joint}(x,y) &= \sum_b w_b\, N_b(x,y), \nonumber\\
V_{\rm joint}(x,y) &= \sum_b w_b^2\, V_b(x,y), \nonumber
\end{align}
\\
and define the joint significance image
\begin{equation}
S_{\rm joint}(x,y) =
\frac{N_{\rm joint}(x,y)}{\sqrt{V_{\rm joint}(x,y)}}.
\end{equation}

Peaks in $S_{\rm joint}$ define the initial source list. This ensures that sources bright in only one band remain detectable while sources present in multiple bands receive a higher combined significance.

The choice of band weights reflects the assumed source spectral energy distribution and can be tuned to specific science goals. For example, in optical surveys targeting dropout populations, weights may be set to zero in blue bands and near unity in red bands to enhance sensitivity to red sources. For typical WISE sources with $W1 - W2 \sim 0$, equal weights across bands are appropriate. For redder or higher-redshift sources in WISE, relatively larger weight on W2 can improve detection significance. In this work, we adopt equal weights as a simple and general choice, and defer a more detailed exploration of optimal band weighting to future work.

The joint significance image also gives a simple expectation for the gain in detection depth. For a source with total flux $F_b$ in band $b$, the detection significance is
\begin{equation}
{\rm S/N}_b = \frac{F_b}{\sigma_b},
\end{equation}
where $\sigma_b$ is the uncertainty on the fitted source flux in that band.

For bright sources with nearly equal flux across bands and approximately zero color, the joint multiband detection significance is
\begin{equation}
({\rm S/N}_{\rm joint})^2 = \sum_b \frac{F_b^2}{\sigma_b^2}
\;\;\Rightarrow\;\;
{\rm S/N}_{\rm joint} = \frac{F}{\sigma_{\rm joint}},
\end{equation}
where the joint variance is given by inverse-variance combination,
\begin{equation}
\frac{1}{\sigma_{\rm joint}^2} = \sum_b \frac{1}{\sigma_b^2}.
\end{equation}

At fixed detection threshold, the limiting flux scales with the noise,
\begin{equation}
F_{\rm lim} \propto \sigma,
\end{equation}
so that a smaller $\sigma_{\rm joint}$ implies a fainter limiting flux.

Magnitudes are related to flux by
\begin{equation}
m = -2.5 \log_{10} F + {\rm const},
\end{equation}
so the gain in limiting magnitude relative to band $b$ is
\begin{equation}
\Delta m_b
=
2.5 \log_{10}
\left(
\frac{\sigma_b}{\sigma_{\rm joint}}
\right)
=
1.25 \log_{10}
\left(
\frac{\sigma_b^2}{\sigma_{\rm joint}^2}
\right).
\label{eq:deltam_exp}
\end{equation}

These relations show that the improvement from multiband detection is set by how the joint variance combines the individual band variances. Since $\sigma_{\rm joint}$ is dominated by the lowest-noise (deepest) band, the gain relative to that band is expected to be small. In contrast, the noisier band benefits significantly from the inclusion of the deeper band, leading to a much larger improvement in its effective depth.

This provides an idealized expectation for the gain from the joint significance image alone. In practice, the realized improvement can differ because the final catalog also depends on blending, centroiding, sky estimation, and post-fit selection cuts such as signal-to-noise thresholds and neighbor-based pruning.

\subsection{Linear System Construction}
\label{sec:matrix_construction}

Our photometric fitting problem is formulated as a linear least-squares system,
\begin{equation}
\mathbf{y} \approx \mathbf{A}\,\mathbf{x},
\end{equation}
where the data vector $\mathbf{y}$ contains image pixel values, the design matrix $\mathbf{A}$ encodes the photometric model evaluated on those pixels, and the parameter vector $\mathbf{x}$ contains the unknown source and background parameters. The system is solved using an iterative sparse least-squares solver (LSQR) \citep{1982ACMTM...8...43P}.

\subsection{Single-band System}
\label{sec:sb}

In the single-band case used by earlier versions of \texttt{crowdsource}, each band is fit independently.

The data vector for band $b$ is
\begin{equation}
\mathbf{y}_b \in \mathbb{R}^{P},
\end{equation}
containing the pixel values of the image in that band.

We consider a local image region containing $P$ pixels, $N$ sources, and $K$ background (sky) basis components. The detailed construction of the design-matrix sub-blocks corresponding to flux, centroid, and sky terms is described in Appendix~\ref{sec:submatrices}.

The parameter vector is
\begin{equation}
\mathbf{x}_b =
\begin{bmatrix}
\text{fluxes}_b\;(N) \\
\text{centroids}_b\;(2N) \\
\text{sky}_b\;(K)
\end{bmatrix}
=
\begin{bmatrix}
F_{1b} \\ \vdots \\ F_{Nb} \\
\delta x_{1b} \\ \delta y_{1b} \\ \vdots \\ \delta x_{Nb} \\ \delta y_{Nb} \\
s_{b,1} \\ \vdots \\ s_{b,K}
\end{bmatrix}
\end{equation}

The full design matrix is constructed by concatenating submatrices encoding the flux, centroid-shift, and sky components,

\begin{equation}
\mathbf{A}_b =
\Big[
\mathbf{A}^{\mathrm{flux}}_b \;\;
\mathbf{A}^{\mathrm{cent}}_b \;\;
\mathbf{A}^{\mathrm{sky}}_b
\Big]
\;\;\in\;\; \mathbb{R}^{P \times (N + 2N + K)}.
\end{equation}

The resulting system,
\begin{equation}
\mathbf{y}_b \approx \mathbf{A}_b\,\mathbf{x}_b,
\end{equation}
is constructed and solved independently for each band.

\subsection{Multiband System}
\label{sec:mb}

In the multiband case, multiple bands are fit jointly with shared source positions.

For two bands (e.g.\ W1 and W2), the data vectors are stacked,
\begin{equation}
\mathbf{y} =
\begin{bmatrix}
\mathbf{y}_{\mathrm{W1}} \\
\mathbf{y}_{\mathrm{W2}}
\end{bmatrix}
\;\;\in\;\; \mathbb{R}^{2P}.
\end{equation}

The parameter vector contains band-dependent fluxes, shared centroid parameters, and band-dependent sky terms,
\begin{equation}
\mathbf{x} =
\begin{bmatrix}
\text{fluxes}_{\mathrm{W1}}\;(N) \\
\text{fluxes}_{\mathrm{W2}}\;(N) \\
\text{shared centroids}\;(2N) \\
\text{sky}_{\mathrm{W1}}\;(K) \\
\text{sky}_{\mathrm{W2}}\;(K)
\end{bmatrix}.
\end{equation}

The full block-structured design matrix is
\begin{equation}
\mathbf{A} =
\begin{bmatrix}
\mathbf{A}^{\mathrm{flux}}_{\mathrm{W1}} & 0 &
\mathbf{A}^{\mathrm{cent}}_{\mathrm{W1}} &
\mathbf{A}^{\mathrm{sky}}_{\mathrm{W1}} & 0 \\
0 & \mathbf{A}^{\mathrm{flux}}_{\mathrm{W2}} &
\mathbf{A}^{\mathrm{cent}}_{\mathrm{W2}} &
0 & \mathbf{A}^{\mathrm{sky}}_{\mathrm{W2}}
\end{bmatrix}.
\end{equation}

Each block above has the same interpretation as in the single-band case, with centroid-related columns corresponding to shared position parameters across bands and flux and sky terms remaining band-specific (see Appendix~\ref{sec:submatrices}).

The joint system,
\begin{equation}
\mathbf{y} \approx \mathbf{A}\,\mathbf{x},
\end{equation}
is then solved using LSQR, analogously to the single-band case.  However, because the design matrix now contains consistent centroids and centroid shifts for each source, this formulation enforces consistent source positions across bands while allowing fluxes and sky models to vary independently per band.

\section{Testing on unWISE coadds}
\label{sec:tests}

We test the multiband photometry algorithm using the publicly available unWISE coadded images \citep{2019PASP..131l4504M} in the COSMOS field \citep{2007ApJS..172....1S}, which provides a well-studied extragalactic region with extensive ancillary data, including deep \textit{Spitzer} IRAC imaging \citep{2007ApJS..172...86S}, and a wide range of source densities. The unWISE coadds combine individual WISE exposures without additional smoothing, preserving the native WISE resolution while enabling deeper imaging through temporal coaddition \citep{2014AJ....147..108L, 2017AJ....154..161M}. This makes them a suitable testbed for evaluating crowded-field photometric algorithms under realistic survey conditions, where source blending, spatially varying backgrounds, non-uniform noise, and scan-pattern-driven coverage variations are all present.

Our tests are performed on unWISE tiles overlapping the COSMOS field (\texttt{1497p015}, \texttt{1497p030}, \texttt{1512p015}, and \texttt{1512p030}), typical of the high Galactic latitude sky. We apply the multiband algorithm to WISE W1 and W2 imaging bands (centered at 3.4 and 4.6~\micron\ respectively), using band-dependent point-spread function (PSF) models and inverse-variance maps appropriate for each band. We adopt equal relative weights for the two bands in the joint fit. The band weights control the relative contribution of each band to the joint fit, while the inverse-variance maps determine the statistical weighting of individual pixels within each band.

For all catalogs, we apply a signal-to-noise cut of
$\mathrm{S/N} = F/\sigma_F > 5$ on each band and perform cross-matching, where required, using a maximum separation of $2^{\prime\prime}$. These selections are applied uniformly to the single-band, multiband, and external reference catalogs to ensure consistent comparisons.  For the sky background, we adopt the default \texttt{crowdsource} sky treatment used for WISE imaging in \cite{2019ApJS..240...30S}.

In the following sections, we evaluate the resulting photometry in terms of flux consistency, astrometric stability, residual structure, and detection behavior, comparing against the corresponding single-band fits.

\subsection{Faint-end Flux Behavior}

\begin{figure*}
\centering
\includegraphics[width=\textwidth]{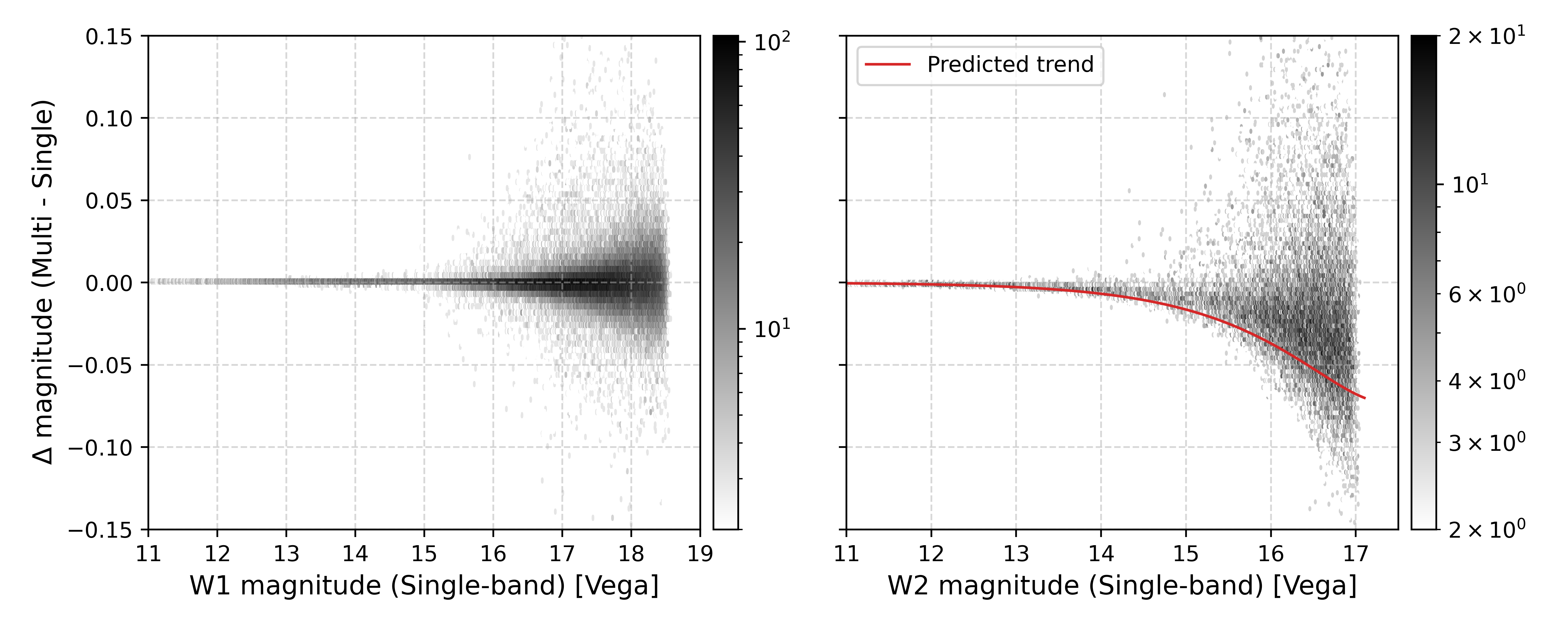}
\caption{Difference between multiband and single-band WISE magnitudes as a function of single-band magnitude for W1 (left) and W2 (right). Bright sources show excellent agreement between the two methods, while systematic differences emerge at the faint end, most prominently in W2. The negative shift in faint W2 magnitudes indicates that the sky level was underestimated in the single-band W2 photometry; including the additional sources detected in the multiband photometry (primarily from W1) we find a fainter sky level and correspondingly mildly more flux in the individual sources (5\% on average at 17 mag).  Joint fitting across bands reduces this sky bias and improves faint-source photometry.
}\label{fig:faint_flux}
\end{figure*}

We first examine the agreement between single-band and multiband flux measurements as a function of magnitude. For bright sources, the two methods agree closely in both W1 and W2. At the faint end, however, W2 fluxes measured in the new multiband W2 catalogs tend to be brighter than the single-band catalogs. Figure~\ref{fig:faint_flux} shows the difference between multiband and single-band magnitudes as a function of single-band magnitude.


The systematic difference between the multiband W2 magnitudes and the single-band magnitudes is small, roughly a quarter of the estimated uncertainty at the faintest magnitudes. At least three effects contribute to the systematic difference: degeneracy between the sky and source fluxes, photometric bias in maximum-likelihood fitting, and blending. Because WISE images are crowded, there is a significant degeneracy between the sky level and the fluxes of faint sources. The new multiband catalogs model many more sources in W2 than are modeled in the single-band catalogs, allowing the multiband fit to attribute more light to sources and less to the sky than was possible in the single-band catalogs. In the single-band W2 fit, some faint-source light is instead absorbed into the sky model, biasing the sky high and making faint source fluxes too low. This leads the multiband W2 fluxes to be brighter than the single-band fluxes, and represents an improvement over the single-band analysis.

Another effect is that maximum-likelihood photometry is biased when the position of a source is unknown, as derived in \citet{2020AJ....159..165P}. In the single-band photometry, the positions of the sources in W2 are free to float to soak up additional noise, leading to a slight overestimation of the fitted flux (i.e., a positive bias). In the multiband case, the positions are largely fixed to the W1 locations, removing this bias and contributing to the difference between the single and multiband fluxes. 

We illustrate the size of these effects with the red curve in Figure~\ref{fig:faint_flux}.  The curve is given by Equation~\ref{eq:deltam_faint},

\begin{equation}
\Delta m_{\rm exp} =
-2.5\log_{10}\!\left[
1 + \frac{\sum p}{\sum p^2} \frac{\,\Delta {\rm sky}}{f_{\rm single}}
-\left(\frac{\sigma_{\rm multi}}{f_{\rm single}}\right)^2\right]
\label{eq:deltam_faint} \, ,
\end{equation}
where $p$ is the PSF, $\Delta \mathrm{sky}$ is the median sky difference between the single-band and multiband models, and $f_\mathrm{single}$ is the single-band flux of a source.  

The sky and bias effects described above correspond to the second and third terms in the logarithm of Equation~\ref{eq:deltam_faint}. Thus, the red curve should be interpreted as a simple model for the mean faint-end behavior.  The most important effect that we have not included in this simple model is that the multiband catalog features much better deblending than the single-band catalog, due to the use of the deeper W1 source catalog, leading to the tail in the distribution of sources toward fainter multiband than single-band fluxes.


\subsection{Color Stability and Bias Reduction}

\begin{figure}
\centering
\includegraphics[width=\columnwidth]{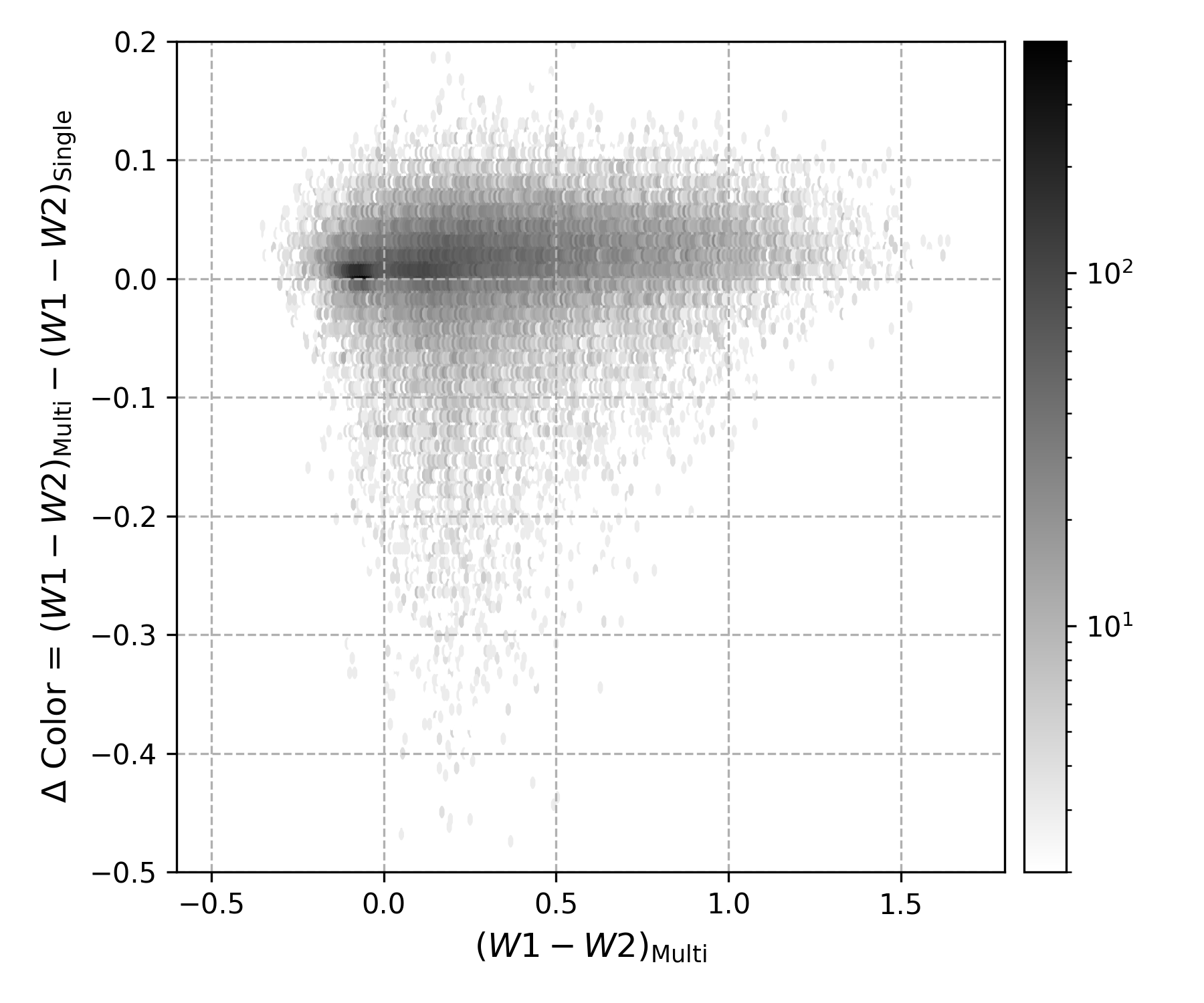}
\caption{Difference in $(W1-W2)$ color between multiband and single-band photometry as a function of the multi-band color. Most sources cluster near zero difference, indicating consistency between the two approaches. Sources with large color differences tend to have average multiband colors, indicating that the multiband colors are more reliable.
}
\label{fig:color_shift}
\end{figure}

To assess the impact of joint fitting on derived colors, we compare $(W1-W2)$ colors measured from single-band and multiband catalogs. Figure \ref{fig:color_shift} shows the difference in color as a function of the multi-band color.

Most sources cluster near zero color difference, indicating consistency between the two approaches. The largest color shifts occur for faint and blended sources, where multiband fitting systematically moves objects toward the main, physically plausible color locus (diagonal plume). This behavior is driven by the use of consistent source locations, particularly in the lower signal-to-noise W2 band.

\subsection{Color Changes in a Crowded Bulge Field}

We also test the multiband algorithm on a representative unWISE Galactic bulge tile (2602m273) as a stress test of its behavior in a more crowded stellar field, where blending is more severe than in the COSMOS field. In this case, we compare the distribution of sources in $(W1-W2)$ color versus W1 magnitude.

\begin{figure}
\centering
\includegraphics[width=\columnwidth]{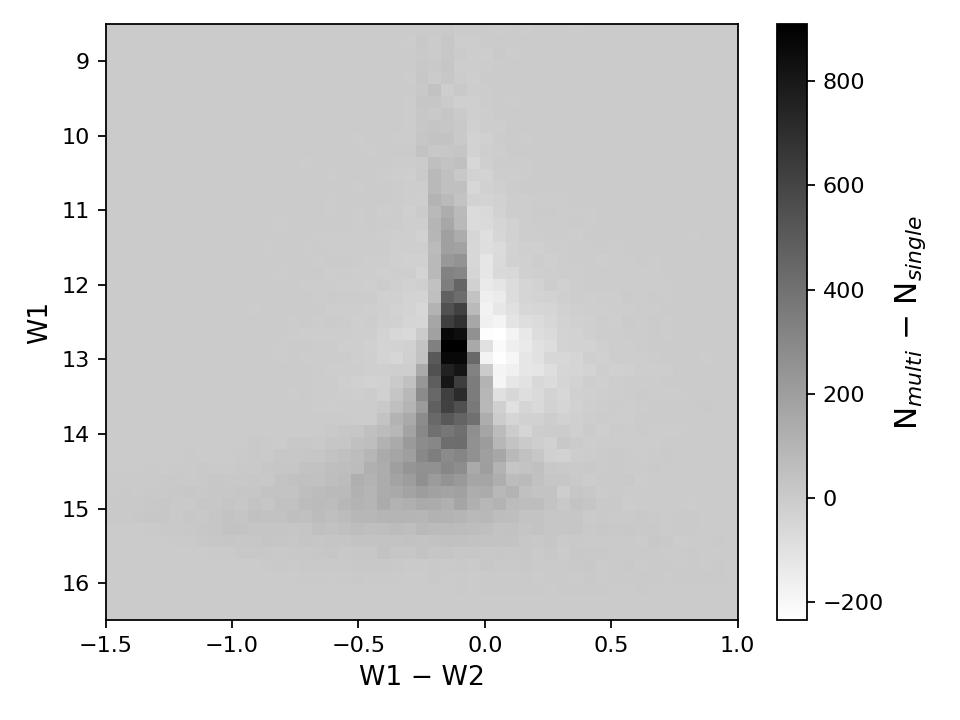}
\caption{
Difference in source counts between the multiband and single-band catalogs in $(W1-W2)$ color versus W1 magnitude for a representative Galactic bulge unWISE field (2602m273). The multiband fit shows a strong excess along the main stellar locus and at the faint end, indicating improved source recovery in crowded conditions.
}
\label{fig:bulge_color_diff}
\end{figure}

Figure~\ref{fig:bulge_color_diff} shows the difference in source counts between the multiband and single-band catalogs across color--magnitude space. Rather than tracing individual source-by-source shifts, this diagnostic highlights where the multiband fit changes the overall distribution of colors.  Figure~\ref{fig:bulge_color_diff} highlights that the multiband catalogs show fewer objects with colors different from the typical star (light), and more sources with typical colors (dark), indicating that the multiband catalog has improved the measured colors of bulge stars.  This behavior is consistent with the interpretation that joint fitting improves source recovery in crowded conditions by using shared positional information across bands, thereby stabilizing photometry in regions where single-band fitting is more affected by blending.

\section{Validation using Spitzer-COSMOS}

To assess completeness and reliability, we compared the resulting unWISE catalogs with the much deeper, higher-resolution mid-infrared Spitzer imaging \citep{2007ApJS..172...86S} on the COSMOS field \citep{2007ApJS..172....1S}. We restrict the analysis to the portion of the tiles covered by the COSMOS field.  The WISE W1 and W2 bands, centered at 3.4 and 4.6~\micron, are close matches to the \textit{Spitzer}/IRAC ch1 and ch2 bands, at 3.6 and 4.5~\micron~respectively. The \textit{Spitzer} data therefore provide higher-resolution, higher-sensitivity imaging enabling external validation of source detection, photometric consistency, and blending behavior.

\subsection{Recovery of Additional Sources}

Joint multiband fitting can recover faint or blended sources that are missed or inconsistently detected in independent single-band photometry. Across the tested tiles, the multiband catalog contains 294,012 sources, compared to 284,474 and 122,773 sources in the corresponding W1 and W2 single-band photometric catalogs.  Relative to W1 alone, the multiband run therefore recovers roughly 10,000 additional detections.

To test the reliability of the newly detected sources, we identify sources detected by the multiband run that lack counterparts in the corresponding single-band WISE catalogs. We conservatively define ``lacking counterparts'' to refer to multiband sources that have no single-band source within $6\arcsec$, in order to focus on new sources that are detected due to the deeper joint significance image rather than new sources that are only deblended differently. Using this criterion, we identify approximately 3850 multiband detections across the tested tiles that have no counterparts in either single-band catalog.

To assess whether these additional detections correspond to real astrophysical objects, we cross-match them against the deeper Spitzer-COSMOS catalogs using a $6\arcsec$ matching radius. Of these $\sim$3,850 new multiband detections, approximately $\sim$3,200 have Spitzer counterparts, indicating that $\sim80\%$ of these faint sources correspond to genuine astrophysical objects.

\subsection{Completeness}
\begin{figure*}
\centering
\includegraphics[width=\textwidth]{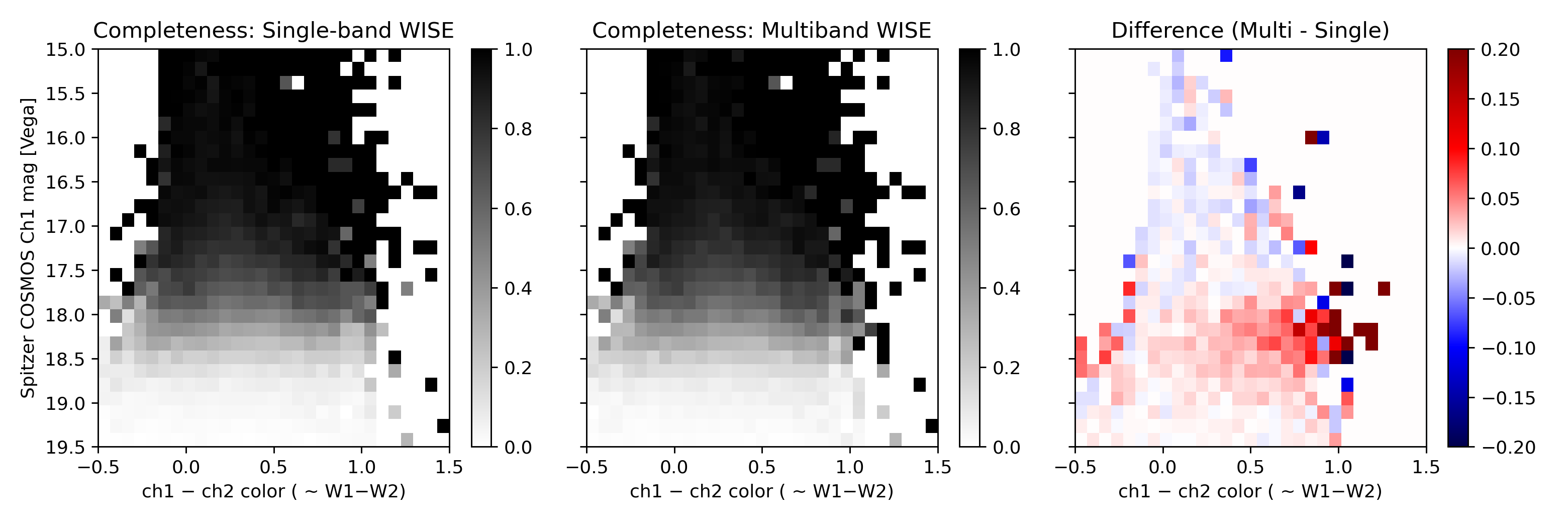}
\caption{ Two-dimensional completeness in the COSMOS field using Spitzer COSMOS as a deeper reference catalog. Left: completeness for single-band WISE photometry. Middle: completeness for multiband WISE photometry. Right: difference between multiband and single-band completeness. Multiband fitting recovers additional sources at faint magnitudes across a range of colors, particularly toward relatively red W1-W2 colors.}
\label{fig:completeness_maps}
\end{figure*}

Figure~\ref{fig:completeness_maps} shows the two-dimensional completeness of the WISE catalogs in the Spitzer-COSMOS field as a function of Spitzer ch1 magnitude and Spitzer color, $\mathrm{ch1}-\mathrm{ch2}$. Following \citet{2019ApJS..240...30S}, for this Figure we use a clean Spitzer reference sample selected by requiring positive flux in both IRAC channels (\texttt{flux\_c1\_4 > 0}, \texttt{flux\_c2\_4 > 0}) and clean quality flags (\texttt{fl\_c1 = 0}, \texttt{fl\_c2 = 0}). We then identify which sources in this cleaned sample are recovered by the single-band and multiband WISE catalogs within $2''$, and compute the recovered fraction in bins of Spitzer magnitude and color. No additional signal-to-noise cut is applied on the WISE side for this analysis.

The left and middle panels show the resulting completeness for the single-band and multiband WISE catalogs, respectively, while the right panel shows their difference. At bright ch1 magnitudes, both catalogs are nearly complete over most of the color range. Toward fainter magnitudes, the completeness decreases as the $S/N$ declines.  The multiband catalog remains more complete at the faint end, demonstrating that joint fitting improves source recovery where the single-band measurements begin to fall off.

The color dependence gives additional physical information. The horizontal axis traces the mid-infrared color of the sources. The difference map shows that the gain from multiband fitting is concentrated mainly at faint magnitudes and is strongest for moderately red sources (lower right of the right panel of Figure~\ref{fig:completeness_maps}). Physically, this means that the joint multiband fit is most effective when both bands contribute significantly to the signal-to-noise.  

We further quantify differential completeness by computing the fraction of Spitzer-COSMOS sources recovered by unWISE as a function of magnitude. The parent Spitzer sample is defined by requiring positive flux in the corresponding IRAC channel and a clean quality flag by setting \texttt{flux\_c?\_4 > 0} and \texttt{fl\_c? = 0} in the relevant channel. One consequence is that the brightest Spitzer sources are underrepresented, so the completeness curves do not extend cleanly to the brightest magnitudes. We further remove Spitzer sources associated with resolved WISE galaxies by identifying image regions with the HyperLeda flag bit~9 set and masking any Spitzer source that falls within those flagged regions. This effectively restricts the completeness analysis to sources significantly smaller than the WISE PSF. Both the Spitzer and unWISE samples are then restricted to the same Spitzer-COSMOS sky region. We also require \texttt{primary = 1} on the unWISE catalogs that indicates whether a particular source is located in the “primary” region of its coadd.

\begin{figure*}
\centering
\includegraphics[width=\textwidth]{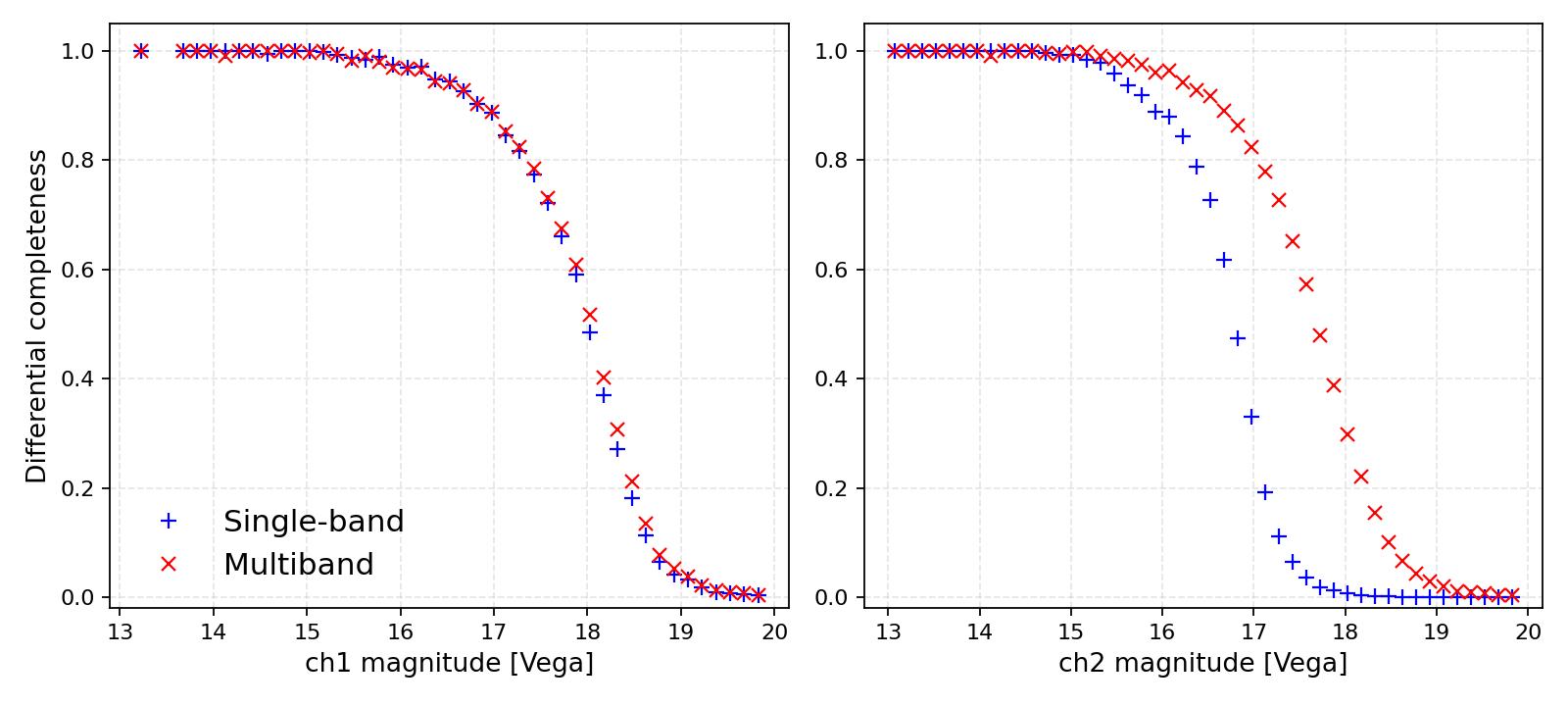}
\caption{Differential completeness as a function of Spitzer magnitude, defined as the fraction of Spitzer-COSMOS sources recovered in WISE per magnitude bin. Single-band (blue) and multiband (red) results are shown for Spitzer ch1 (left) and ch2 (right). Both methods achieve near-unity completeness at bright magnitudes, while multiband fitting remains complete to fainter limits. The largest improvement is observed in ch2/W2, where single-band completeness declines earlier due to the relative lack of sensitivity of the W2 band for typical sources.}
\label{fig:diff_complete}
\end{figure*}

The differential completeness in each magnitude bin is then measured as the fraction of sources in the cleaned Spitzer sample that have a WISE counterpart within $2''$. Figure~\ref{fig:diff_complete} shows the resulting curves for the single-band and multiband catalogs. At bright magnitudes both methods are nearly complete, while toward fainter magnitudes the multiband run remains complete to fainter limits. The gain is modest in ch1/W1 but much larger in ch2/W2, where the single-band catalog degrades earlier because of the lower sensitivity in W2.

This behavior is consistent with the expectations from Section~\ref{sec:sigimage}, where we derived the expected gain in limiting magnitude for zero-color sources in Equation \ref{eq:deltam_exp}. For the unWISE data considered here, this predicts a substantial improvement in the noisier band (W2), of order $\sim$1.5 mag, and only a negligible gain of $\sim$0.03 mag in the deeper band (W1), as $\sigma_{\rm joint}$ is dominated by the lowest-noise band. In Figure~\ref{fig:diff_complete}, we observe an improvement of nearly $\sim$1 mag in W2, broadly consistent with this expectation. The modest difference reflects the idealized assumptions in the derivation, in particular the assumption that sources have zero color (for example, for red galaxies with W1$-$W2 = 1, the expected gain is 0.4). Overall, the observed gain follows the expected trend, with minimal improvement in W1 and a larger gain in W2.

\subsection{Differential Reliability}

Finally, we assess differential reliability, defined as the fraction of WISE sources that have a Spitzer-COSMOS counterpart. For the Spitzer matching sample, we keep sources with positive flux in the relevant IRAC channel. We do not apply the Spitzer quality cuts used in the completeness analysis, since for reliability statistics we want to compare against a complete sample, in order to avoid penalizing WISE for detecting poorly-measured Spitzer sources.  For the unWISE single-band and multiband catalogs, we again restrict the catalogs to the Spitzer-COSMOS region and require \texttt{primary==1} and $S/N > 5$ in the relevant WISE channel. We remove sources with \texttt{flags\_unwise \(\neq 0\)} and reject sources with any of \texttt{flags\_info} bits 1, 6, or 7 set, since these bits identify problematic or contaminated detections that should not contribute to a purity estimate. 

\begin{figure*}
\centering
\includegraphics[width=\textwidth]{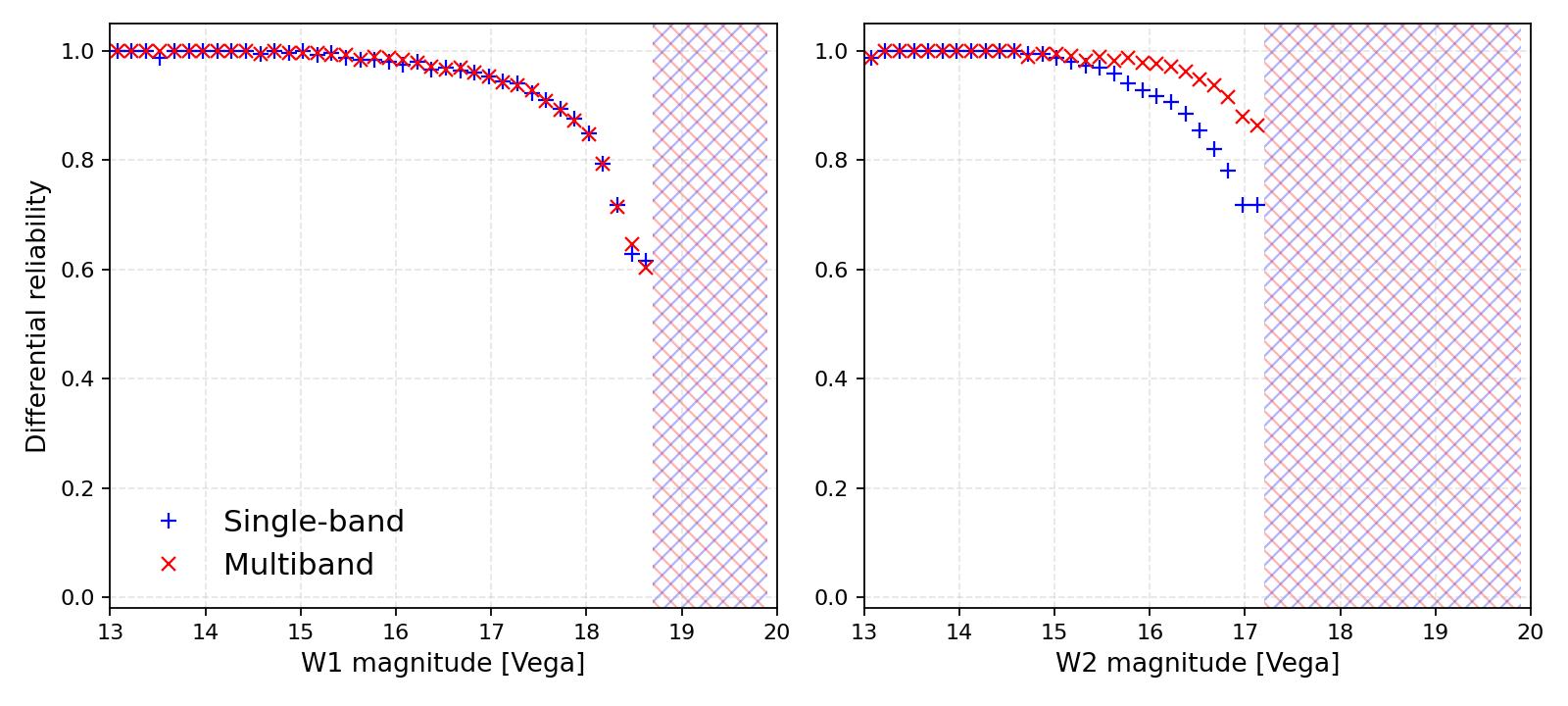}
\caption{Differential reliability (purity) as a function of WISE magnitude, defined as the fraction of WISE sources in each magnitude bin that have a Spitzer-COSMOS counterpart. Single-band (blue) and multiband (red) results are shown for W1 (left) and W2 (right). Multiband fitting maintains high reliability while extending to fainter magnitudes, with the most significant improvement observed in W2 demonstrating that joint fitting suppresses spurious detections without sacrificing completeness. The hatched regions indicate bins with fewer than 10 WISE sources, where the measurements are dominated by small-number statistics.}
\label{fig:diff_reliable}
\end{figure*}

The differential reliability is then measured as the fraction of WISE sources in each magnitude bin that have a Spitzer counterpart within $2''$. Figure~\ref{fig:diff_reliable} shows the resulting curves. Multiband fitting maintains high reliability while extending to fainter limits, with the strongest improvement again appearing in W2. This indicates that joint fitting suppresses spurious detections while still preserving the gain in depth. At the faint end, bins with fewer than 10 WISE sources are marked separately in the figure, since those bins are dominated by small-number statistics. 


\section{Conclusion}
\label{sec:conclusion}

We have presented a multiband extension of the \texttt{crowdsource} photometric pipeline that enables simultaneously modeling multiple images with a consistent set of sources. The algorithm retains the core structure and computational efficiency of the original single-band framework, while incorporating band-dependent point-spread functions, noise models, and flux parameters within a unified linearized least-squares formulation. Source detection is performed using a multiband significance image that combines information across bands while accounting for their individual PSFs and noise properties. Tests on the unWISE coadds in the COSMOS field \citep{2019PASP..131l4504M}  demonstrate that the multiband approach yields more stable centroid estimates, improved flux consistency between bands, and the recovery of additional faint sources, including objects that are individually marginal in both single-band catalogs.

Our main results can be summarized as follows:
\begin{itemize}
    \item We have generalized the \texttt{crowdsource} algorithm to jointly fit multiple imaging bands with a shared set of source positions.
    \item We have developed and implemented a multiband generalization of the significance image that improves source detection by combining information across bands in a statistically consistent way to deliver the deepest possible detections of sources of a known spectral energy distribution.
    \item We have demonstrated the improved performance of the pipeline in terms of source colors, completeness, and reliability compared to independent single-band fits.
\end{itemize}

In this work, we have focused on presenting the generalizations of the \texttt{crowdsource} pipeline to multiple images and validating the success of those extensions in controlled test regions.  In future work we intend to apply this multiband algorithm to the full set of unWISE coadds across all WISE and NEOWISE imaging, to construct the deepest possible catalog of WISE sources to date. Such a catalog will be valuable for many of the same applications already enabled by the unWISE Catalog, including Galactic studies, extragalactic galaxy samples, and high-redshift quasar searches, as well as cosmological analyses based on unWISE-selected galaxies such as CMB lensing tomography \citep{2020JCAP...05..047K, 2021JCAP...12..028K, 2025PhRvD.111h3516F}.

The multiband framework also enables flexibility in the relative weighting of different bands in the joint fit. In this paper, we have adopted equal relative band weights, allowing the pixel-level inverse-variance maps to determine the effective statistical weighting. Future work will explore alternative band-weight choices like emphasizing the W2 band to enhance sensitivity to higher-redshift sources. Varying the relative band weights and examining the resulting detection significance and catalog properties provides a practical way to optimize source recovery for different science goals.

Beyond static imaging, this framework can be naturally extended to incorporate multi-epoch data, where repeated observations can be combined within the same modeling framework. In the simplest case, this improves sensitivity by increasing the effective signal-to-noise, while more general extensions could enable modeling of time-variable and transient sources. The multiband \texttt{crowdsource} framework provides a natural foundation for these extensions and for future improvements in source detection and photometry in crowded, multi-epoch survey data.

\begin{acknowledgments}
J.B. and E.S. acknowledge support from NASA ADAP grant No. 80NSSC25K7562.

We acknowledge the use of the HyperLeda database (\url{http://leda.univ-lyon1.fr}).

This publication makes use of data products from the Wide-field Infrared Survey Explorer, which was a joint project of the University of California, Los Angeles, and the Jet Propulsion Laboratory/California Institute of Technology, and from NEOWISE, which was a project of the Jet Propulsion Laboratory/California Institute of Technology. WISE and NEOWISE are funded by the National Aeronautics and Space Administration.

\end{acknowledgments}

%
\facility{WISE}
\software{astropy \citep{2013A&A...558A..33A} }


\appendix

\section{Details of the \texttt{crowdsource} pipeline}
\label{sec:appendix_pipeline}

This section provides implementation details for each step of the \texttt{crowdsource} pipeline outlined in Section \ref{sec:algo}.

\subsection{Smooth Sky Subtraction}
\label{sec:appendix_sky}

We begin each iteration by subtracting a smooth background from the image. This background is estimated using a median-filtered version of the current residual image, which captures large-scale structure while limiting the influence of point sources. This step aims to isolate the astronomical point sources of interest from detector and sky backgrounds.  As the process continues, point sources are subtracted increasingly well and become less and less important.

A second sky component is optionally included during the linear fitting (see Section~\ref{sec:appendix_fitting}). Thus, the full sky model is a combination of (i) a median filtered image and (ii) a model-based sky component.

\subsection{Source Detection}
\label{sec:appendix_detection}

Source detection is performed on the residual image after smooth sky subtraction. Candidate sources are identified as peaks in the matched-filter significance image described in Section~\ref{sec:sigimage}.


Newly detected peaks are added to the model at each iteration. To avoid overfitting noise, detections must exceed a significance threshold, and sources that fail to maintain sufficient signal-to-noise in subsequent iterations are removed (see Section~\ref{sec:appendix_pruning}).

\subsection{Linear Fit: Fluxes, Positions, and Sky}
\label{sec:appendix_fitting}

Given the current source list, we solve for source fluxes, position corrections, and sky parameters using the linear system described in Section \ref{sec:matrix_construction}.

The system is constructed using band-dependent PSFs and inverse-variance maps. Only pixels within the support of each source’s PSF are included, leading to a sparse design matrix.

The matrix is solved using the LSQR algorithm, which efficiently handles large sparse systems. In practice:
\begin{itemize}
    \item The data vector is weighted by inverse variance.
    \item The design matrix includes flux, centroid, and sky sub-blocks
    (Appendix \ref{sec:submatrices}).
    \item Position terms are linearized using PSF derivatives.
\end{itemize}

In addition to source parameters, the fit includes a sky component, modeled using low-order basis functions (e.g., constant or gradients). This model-based sky is solved simultaneously with source parameters and captures residual large-scale structure not removed by the initial median subtraction.

In the multiband case, fluxes are solved independently for each band, while centroid parameters are shared across bands, enforcing a common source position.  Future extensions could allow more flexible modeling, e.g., allowing for source motion and parallax.

\subsection{Position Refinement}
\label{sec:appendix_centroid}

Source positions are updated iteratively using centroid corrections obtained from the linear fit. In addition, centroid estimates are computed directly from neighbor-subtracted images, where contributions from nearby sources are removed using the current model. These measurements provide robust position updates, particularly in crowded regions.

In the multiband formulation, centroid updates are driven by the combined signal across all bands, improving positional stability when one band is noisier or shallower.

\subsection{PSF Refinement}
\label{sec:appendix_psf}

The PSF model is refined using residuals between the data and the current source model. Bright, isolated sources are used to update the PSF, improving its accuracy across the field. This refinement is performed iteratively and allows the PSF model to adapt to spatial variations and mismatches in the initial PSF estimate.





\subsection{Source Pruning and Heuristics}
\label{sec:appendix_pruning}

The \texttt{crowdsource} pipeline applies a set of heuristic criteria during both source detection and iterative fitting to suppress noise fluctuations, avoid over-deblending, and maintain a stable source list.

At the detection stage, candidate sources are identified as local maxima in the significance image with ${\rm S/N} > 5$, corresponding to isolated sources in the presence of photon and detector noise. Peaks that are too strongly blended with the current model are rejected using a blending threshold $B$ (set to $B = 0.3$ in this work), based on comparisons between the image and model as well as their corresponding significance images. Specifically, peaks are retained only if one of the following conditions is satisfied:

\begin{enumerate}
    \item $\mathrm{S_I} / \mathrm{S_M} > 2B$, where $\mathrm{S_I}$ is the significance image and $\mathrm{S_M}$ is the model significance image, or
    \item $\mathrm{S_I} / \mathrm{S_M} > B$ and $(I/M) > B$, where $I$ and $M$ denote the image and model values (without smoothing to construct the significance image) at the peak location.
\end{enumerate}

These criteria suppress detections in regions already well described by the model, preventing the addition of spurious or over-split components.

After each fitting iteration, additional pruning is applied to remove low-significance and unstable sources. Sources are required to exceed a minimum signal-to-noise threshold of $\sim 3\sigma$. In addition, lower-significance sources within 1 pixel of a higher-significance neighbor are removed, suppressing duplicate detections and unstable subcomponents in crowded regions.

Together, these steps maintain a stable source list by ensuring that the model remains focused on statistically significant and well-isolated sources while allowing progressively fainter sources to be incorporated over successive iterations.





\section{Sub-matrices}
\label{sec:submatrices}

In this appendix, we discuss the explicit structure of the design matrix sub-blocks used in the linearized least-squares formulation described in Section~\ref{sec:matrix_construction}.

For a given band $b$, we decompose the design matrix into three conceptual blocks, corresponding to flux, centroid, and sky terms.

\subsection{Flux submatrix}

The flux submatrix,
\begin{equation}
\mathbf{A}^{\mathrm{flux}}_b \; (P \times N) \;\approx\;
\begin{bmatrix}
P_{1b} & P_{2b} & 0      & \cdots & 0 \\
P_{1b} & P_{2b} & 0      & \cdots & 0 \\
P_{1b} & P_{2b} & P_{3b} & \cdots & 0 \\
0      & P_{2b} & P_{3b} & \cdots & 0 \\
0      & 0      & P_{3b} & P_{4b} & \vdots \\
\vdots & \vdots & \vdots & \ddots & P_{Nb}
\end{bmatrix}
\end{equation}
contains one column per source. Here $P$ is the number of pixels in the fitting region and $N$ is the number of sources included in the local fit. Each column represents the point-spread function (PSF) of a source evaluated over all pixels in the fitting region. This block encodes how each source contributes flux to the image pixels. Pixels far from any source have zero, or negligible, contribution, so this matrix is sparse in practice.

\subsection{Centroid submatrix}

The centroid submatrix,
\begin{equation}
\mathbf{A}^{\mathrm{cent}}_b \; (P \times 2N) \;\approx\;
\begin{bmatrix}
D^x_{1b} & D^y_{1b} & D^x_{2b} & D^y_{2b} & 0        & 0        & \cdots \\
D^x_{1b} & D^y_{1b} & D^x_{2b} & D^y_{2b} & 0        & 0        & \cdots \\
D^x_{1b} & D^y_{1b} & D^x_{2b} & D^y_{2b} & D^x_{3b}& D^y_{3b}& \cdots \\
0        & 0        & D^x_{2b} & D^y_{2b} & D^x_{3b}& D^y_{3b}& \cdots \\
0        & 0        & 0        & 0        & D^x_{3b}& D^y_{3b}& \cdots \\
\vdots   & \vdots   & \vdots   & \vdots   & \vdots  & \vdots  & \ddots
\end{bmatrix}
\end{equation}

where
\begin{equation}
D^x_{ib} \equiv F_{ib}\,\partial_x \mathrm{PSF}_{ib},\qquad
D^y_{ib} \equiv F_{ib}\,\partial_y \mathrm{PSF}_{ib}.
\end{equation}


contains two columns per source, corresponding to the response of the model to small shifts in the source position along the $x$ and $y$ directions. Conceptually, these columns are PSF-derivative templates and encode how pixel values change when a source centroid is perturbed. Here $F_{ib}$ denotes the current estimate of the source flux, taken from the previous iteration of the fit. For newly detected sources in the first iteration, a unit flux is assumed, and the resulting centroid parameters are subsequently rescaled by the current fitted flux.



\subsection{Sky submatrix}

The sky submatrix for band $b$ contains the sky basis functions evaluated at each pixel in the fitting region,
\begin{equation}
\mathbf{A}^{\mathrm{sky}}_b \; (P \times K) \;\approx\;
\begin{bmatrix}
B_{1b} & B_{2b} & \cdots & B_{Kb} \\
B_{1b} & B_{2b} & \cdots & B_{Kb} \\
B_{1b} & B_{2b} & \cdots & B_{Kb} \\
\vdots & \vdots & \ddots & \vdots \\
B_{1b} & B_{2b} & \cdots & B_{Kb}
\end{bmatrix}
\end{equation}

Here $P$ is the number of pixels and $K = n_{\mathrm{sky},x} \times n_{\mathrm{sky},y}$ is the number of sky parameters. The sky is modeled locally within each fitting subregion and solved simultaneously with the source fluxes. This block models large-scale sky structure within the fitting region.

\bibliography{my}{}
\bibliographystyle{aasjournalv7}

\end{document}